\documentclass[sigconf]{acmart}

\usepackage[english]{babel}
\usepackage{blindtext}

\renewcommand\footnotetextcopyrightpermission[1]{} 
\setcopyright{none}

\acmDOI{}

\acmISBN{}

\acmPrice{}

\usepackage{balance}

\usepackage{subfigure}
\usepackage{graphicx}
\usepackage{amssymb}
\usepackage{amsmath}
\usepackage{multicol,multirow}
\usepackage{epsf,psfrag,pstricks}
\usepackage{mdwlist}
\usepackage{rotating}
\usepackage{xspace}
\usepackage{url}
\usepackage{capt-of}
\usepackage{flushend}

\makeatletter
\renewcommand{\paragraph}[1]{\vspace*{0.05in}\noindent{\bf #1}\hspace{0.25ex \@plus1ex \@minus.2ex}}
\makeatother

\begin{document}

\renewcommand{\sectionautorefname}{\S}
\renewcommand{\subsectionautorefname}{\S}

\title[]{The Dagstuhl Beginners Guide to Reproducibility for Experimental Networking Research} 

\author{Vaibhav Bajpai}
\affiliation{%
  \institution{TU Munich}
}
\email{bajpaiv@in.tum.de}

\author{Anna Brunstrom}
\affiliation{%
  \institution{Karlstad University}
}
\email{anna.brunstrom@kau.se}

\author{Anja Feldmann}
\affiliation{%
  \institution{MPI for Informatics}
}
\email{anja@mpi-inf.mpg.de}

\author{Wolfgang Kellerer}
\affiliation{%
  \institution{TU Munich}
}
\email{wolfgang.kellerer@tum.de}

\author{Aiko Pras}
\affiliation{%
  \institution{University of Twente}
}
\email{a.pras@utwente.nl}

\author{Henning Schulzrinne}
\affiliation{%
  \institution{Columbia University}
}
\email{hgs@cs.columbia.edu}

\author{Georgios Smaragdakis}
\affiliation{%
  \institution{TU Berlin}
}
\email{georgios@inet.tu-berlin.de}

\author{Matthias W\"ahlisch}
\affiliation{%
  \institution{Freie Universit\"at Berlin}
}
\email{m.waehlisch@fu-berlin.de}

\author{Klaus Wehrle}
\affiliation{%
  \institution{RWTH Aachen University}
}
\email{klaus@comsys.rwth-aachen.de}

\renewcommand{\shortauthors}{} 


\author{}
\affiliation{
\begin{tabular}{c}
\hspace{-2.35in}{\normalsize This article is an editorial note submitted to CCR. It has NOT been peer reviewed.}\\
\hspace{-2.35in}{\normalsize The authors take full responsibility for this article's
technical content. Comments can be posted through CCR Online.}
\end{tabular}
}

\begin{abstract}
    Reproducibility is one of the key characteristics of good science, but hard to achieve for experimental disciplines like Internet measurements and networked systems. This guide provides advice to researchers, particularly those new to the field, on designing experiments so that their work is more likely to be reproducible and to serve as a foundation for follow-on work by others.
\end{abstract}

\maketitle

\vspace{-3em}
\section{Introduction}\label{sec:introduction}

\label{sec:audience}

Good scientific practice makes it easy for researchers other than the authors to reproduce, evaluate and build on the work. Achieving these goals, however, is often challenging and requires planning and care. We attempt to provide guidelines for researchers early in their career and students working in the field of experimental networking research, and as a reminder for others. We begin by summarizing the terminology (\autoref{ACM-terminology}) that will be used throughout this article. We then elaborate the goals and principles (\autoref{sec:goals-principles}), describe best practices required for reproducibility in general (\autoref{sec:best-practices}) and for specific research methodologies (\autoref{sec:best-practises-by-area}), provide tool recommendations (\autoref{sec:tool-recommendations}) and point to additional resources~(\autoref{sec:resources}).

\subsection{ACM Terminology}\label{ACM-terminology}

The terms repeatability, replicability and reproducibility are often used interchangeably and may not necessarily be used consistently within or across technical communities. Since the Association for Computing Machinery (ACM)~\cite{ACM-Artifact} publishes a significant fraction of papers in networked systems and Internet measurements, we draw on their definitions and summarize them in Table~\ref{tab:terms}.

\begin{table}[t]
    \caption{Repeatability, replicability, and reproducibility as defined by ACM~\cite{ACM-Artifact}.}
    \label{tab:terms}
    \centering
    \begin{tabular}{lll}
      \toprule
        & \multicolumn{2}{c}{Level of change} \\ \cmidrule(r){2-3}
        Term  &  Team & Setup \\ \midrule
         \emph{Repeatability}  &  same & same \\
         \emph{Replicability}   &  different & same \\
         \emph{Reproducibility}   &  different & different \\
      \bottomrule
    \end{tabular}
\end{table}

\textbf{Repeatability} is achieved when a researcher can obtain the same results for her own experiment under exactly the same conditions, i.e., she can reliably repeat her own experiment (``Same team, same experimental setup'').

\textbf{Replicability} allows a different researcher to obtain the same results for an experiment under exactly the same conditions and using exactly the same artifacts, i.e., another independent researcher can reliably repeat an experiment of someone other than herself (``Different team, same experimental setup'').

\textbf{Reproducibility} enables researcher other than the authors to obtain the same results for an experiment under different conditions and using her self-developed artifacts (``Different team, different experimental setup'').

\subsection{Goals and Principles}\label{sec:goals-principles}

One of the fundamental hallmarks of science is that research results produced by one team can be replicated or reproduced (\autoref{ACM-terminology}) by another team. Ideally, the second team should only need their general knowledge of the discipline and the details provided in the published paper, complemented by auxiliary materials such as software documentation or technical reports in some cases.

However, repeatability, replicability, and reproducibility are about more than just following the scientific method and being a ``good research citizen''. By carefully documenting workflow and following best practices, other team members in your research group can continue earlier work and build on it. Often, you yourself will need to revisit earlier work, e.g., when compiling your research for your dissertation or a journal paper, recreating results or updating them to reflect new related work or changes in the environment. Nobody likes spending time on reverse-engineering your own code written a year ago or code written by somebody else, investigating why software packages do not compile or wondering whether you can trust the experimental data you gathered. Besides facilitating progress in science, following best practices will also make mistakes less likely or at least easier to find.

The practices described below work best if followed early on, not just as the final step when completing a project.
\section{General Best Practices}\label{sec:best-practices}

Long before you write a paper, the following best practices help to ensure that your research will succeed and that you can trust
your results. 

\subsection{Problem Formulation and Design}\label{sec:evaluation-setup-execution}


\begin{description}
\item[Hypothesize:] ``Think first, run later'': Formulate and document your hypothesis, design the experiments to validate (or not) the hypothesis, conduct the necessary experiments, and finally check the hypothesis. Indeed, often the outcome of an experiment should lead you to revisit the hypothesis. But sometimes, if an experiment does not give you the predicted results or gives you results that seem a little too good to be true, this may  be due to a mistake in the analysis chain. Therefore, each step needs to be validated and cross-checked. As such it is good practice to double check results with others who may be able to spot problems, e.g., your advisor, someone from the organization responsible for the infrastructure on which the data was gathered, or the author of a software component you used. If you work in a small team, it is a good idea to plan the work so that different persons work on different results so that each one can cross check the work of the others.

\item[Plan and solicit early feedback:] Plan and prototype how you want to present your results as early as possible. Visualizations are necessary to explain your results, but they also help you spot anomalies. You should be able to explain notches, spikes or gaps in your graph by something beyond randomness. Follow guidelines for exploring the parameter space, e.g., an ANOVA experimental design. Get feedback early and often: before you start your project, after your initial experimental design, after your first small-scale results, and after your first large-scale results.

\item[Iterate:] You will likely end up having to redo steps as you modify the system under test or improve your measurements and data analysis scripts. Record steps and automate them, e.g., in scripts or Makefiles, so that you are less likely to forget to set a command line parameter, for example. How often do you need to repeat your measurements to eliminate transient factors and gain confidence? Especially when measuring operational systems such as data centers or the Internet, one-time measurements are prone to be biased by transient effects, temporary congestion or just the particular time of day. Those factors should be accounted for when actually planning the measurement.

\item[Factor dynamism:] Generally expect that operational systems you are measuring against are not static during your measurements. There is evidence that well-known Internet services change constantly and that there are ongoing experiments run by service providers that may interfere with your own measurements.
\end{description}

\subsection{Documentation}\label{sec:documentation}

\begin{description}
\item[Record the experiment:] Documenting \emph{all} steps and observations is critical. Scientists in the natural sciences keep lab notebooks for a reason --- follow their example. The lab notebook can be an electronic shared document, recording each step and each resulting observation. Record mistakes, too, so that others do not have to repeat them. If the lab notebook is electronic, recording script executions can be a first step to automating the workflow. It is often tempting to skip documenting code until later when there is supposedly more time, but that time never seems to occur. Research artifacts often live longer than you anticipate and may be shared with other members of the research team. Thus, code as if you are your colleague who has to pick up your project.

\item[Treat metadata as data:] Any data file or database needs to be accompanied by metadata to help you and others understand how the data was created, what it contains, where to find its documentation, and how to recreate it. Metadata can be conveyed via file naming, contained in header sections in the data, or stored separately in a data log that references file names and, to avoid accidental file name reuse, file hashes. Consider automating the generation of the ``mechanical'' metadata in the scripts or tools you write, preferably in some machine-readable format such as JSON or XML.

\item[Use a version control system:] Using a version control system for code, documentation, paper text, as well as experimental results is essential. This will help you determine if a change in measured results might be due to an innocent-looking code change and which experiments you might need to run again. Whenever possible, you create a release of your own software that you used to create the publishable results. Note that including the raw experimental data may or may not be feasible due to size, privacy, or other constraints.

\item[Keep regular backups:] Keep backups. There is nothing more upsetting than losing the original data of a paper that you are about to publish or that already got published. This also avoids digging into the file systems of graduate students who have long left the university and hoping that their account has not been deleted. Indeed, the data management plans for most organizations and research grants require that scientific artifacts are not only documented but also preserved for multiple years (e.g., five to ten years). Most research institutions offer resources to store data safely and with flexible access control policies.

\end{description}

\subsection{Experimentation and Data Collection}\label{sec:evaluation-measurement}

\begin{description}
\item[Validate and scale:] Start small and then expand. Run small sample sets, where you can readily predict the results, to understand and verify your tools, approaches and analysis setup. These can then later be used as test cases and sanity checks to ensure that the analysis pipeline is still working even if one of the components gets updated. Use a tool chain to first validate previously published results to ensure that there are no fundamental flaws in the analysis or your understanding of the problem. A welcome side effect is that this often leads to  insights which lead to new research results.

\item[Do not reinvent the wheel:] Before initiating a major software development project check if there is a tool that solves your problem. Creating your own tool may bring you to face issues that others have already solved. More than that, creating your own tool also likely commits you to maintaining it. Think about convenient ways of decomposing your problem to follow the Unix philosophy of building simple, modular, and extensible code that can be easily maintained, tested and re-purposed.

\item[Monitor your experiment:] Make sure to monitor your tool chain, preferably by automated checking tools. Common problems include running out of disk space and, therefore, creating zero-length files; reboot of a machine without restarting the tools or causing log files to be overwritten; wrong permissions, e.g., when access tokens time out; network failures and, therefore, missing results from a remote machine or API and finally, resource leaks, such as too many open files, that prevent or distort data gathering.
\end{description}

\subsection{Handling Data}\label{sec:data-privacy-integrity-copyright}

\begin{description}

\item[Data privacy, data anonymization and ethics:] Most datasets have privacy constraints that you need to respect. You should never try to de-anonymize data, as that is unethical and will likely discourage others from making data available. Before making data available to others, consider whether it raises any privacy concerns and whether these concerns can be alleviated by anonymization. If in doubt, always consult other members of your research team, more senior researchers, local ethics panel or institutional review board (IRB) and refer to published community guidelines~\cite{Menlo-report, Menlo-report-companion} on ethical principles guiding scientific research. Data that may seem unlinkable by itself can now often be de-anonymized by drawing on external data sources.

\item[Data integrity:] Check for the integrity of your data and account for observation biases. Did you consider synchronization between system elements, randomization, the effects of caching? When evaluating the performance of a system, will likely use cases depend on the average, best or worst-case performance or some ``likely'' worst-case performance?

\item[Licensing and giving credit:] Consider early how the code you use or write will be licensed. Can you share copyrighted code that you purchased or have access to through your institution with your team or the public? Does everyone on your team agree with how you intend to license code you wrote? (For instance, your role in the institution may determine whether your code is for-hire work or your own.) Does the code license require you to make modifications publicly available? Do code or data use Creative Commons~\cite{CC} or open source licenses~\cite{opensource-licenses} that mandate giving credit to sources? Does your research institution or the organization providing research support have guidelines you need to be aware of? For example, some research funding agencies strongly encourage giving credit to their funding, using template text. Consider that often the most restrictive software license for a system determines whether others can use it. But even restrictive code licenses do not prevent sharing of output data or results.
\end{description}
\section{What Should Be Documented?}\label{sec:best-practises-by-area}

Each paper or thesis should document key experimental conditions, possibly in an appendix or separate technical report for lengthy descriptions of details. Many of these experimental conditions that are needed to make your work reproducible are similar for all basic types of experimental networking research, often used in combination: simulation (\autoref{sec:simulation}), prototyping (\autoref{sec:software-implementation-performance-prototyping}), network measurements (\autoref{sec:measurements-real-systems}) and human factors experiments (\autoref{sec:human-subject-experiments}). We describe considerations for each methodology in turn below.

\subsection{Simulations}\label{sec:simulation}

Simulation is a well-known method to understand and validate a proposed concept, protocol or a system. When simulating a system under test (SuT), a model of this SuT is used and its behavior under varying input and configurations analyzed. Your analysis depends completely on the chosen model and will only reflect the characteristics of the model. Therefore, choose your model with care -- whether you create it yourself or use the model somebody else created. Furthermore, consider the granularity at which you plan to simulate, such as traffic flows, individual packets or the physical channel model. Ultimately, being aware of the strategies~\cite{floyd:ton:2001} for accommodating the difficulties in simulation the Internet due to its immense heterogeneity and dynamism is crucial for sound scientific research.

In order for someone to \emph{repeat} your simulation results, your simulation code and input data should be well packaged and documented such that someone can easily re-run your simulation, e.g., by just executing a Makefile or script. In order to be able to \emph{reproduce or replicate} your results, other researchers should also understand why you chose the particular simulation parameters.

\begin{description}
\item[Software setup:] Describe the simulation software, including the version and required run-time environment. Which additional tools such as traffic generators, topology models, analysis tools are required? Which versions were used? Does your simulation require any specific run-time or execution environment, such as many cores or massive amounts of RAM, that may exceed what is commonly available?

\item[Data input and configuration:] Describe the network or system topology including transmission rate, bit error rates, and propagation delays. What traffic traces or models did you use? What were the parameters of the models, including units? (Be particularly careful with easily confused units, such as kb/s (kilobits [1000 bits] per second) vs. KB/s (Kilobytes (1024 bytes) per second).) If you are including a model of the physical channel, such as a wireless link, what parameters did you choose and are they meant to represent a particular real-world environment? If aspects of your traffic or system parameters are chosen randomly, describe which and how you generated the random variables. If random number generator seeds matter, provide them.  Any simulator configuration file that can be shared?

\item[Limitations:] Is your simulation limited in some important way, e.g., in terms of scale or the execution time needed? How does your simulation abstract and simplify the system you are modeling?

\item[Experiments:] How often did you repeat the experiment and how did you choose the repeat count? How did you initialize the system, e.g., were caches cleared before each run? How did you space your parameters? Did they cover the desired design space for your system?

\item[Analysis:] In general, data is sacrosanct and all raw data should be archived. How did you prepare the data? Did you remove any outliers or obvious measurement errors? Did outages or errors leave gaps in your data gathering? How are you accounting for start-up and transient effects? Were there any anomalies? How are you showing the strength of your evidence, e.g., by confidence intervals, variance, ANOVA, goodness-of-fit testing? How did you choose the parameters for statistical tests? Did you change your measurement approach to, for example, meet a $p$-value or confidence interval threshold? If you are testing a hypothesis, how strong is the evidence that the results are not due to random chance?

\item[Presentation:] Did you include all units for all axes in a clear and unambiguous way? Captions for plots should explain the setting and contain all major parameters so that the caption and figure can stand alone. Consider data formats that allow including the plot points or complement plots with tables showing raw data in an appendix or an extended technical report.

\item[Data access:] If your simulation depends on input data other than parameterized random variables, such as traces or topologies, these should be included with the simulation code or stored in a publicly accessible repository -- see \autoref{sec:software-implementation-performance-prototyping}.
\end{description}

\subsection{Systems Prototyping and Evaluations}\label{sec:software-implementation-performance-prototyping}

To evaluate a new protocol, service or algorithm you can build a prototype and then measure its scalability, performance or efficiency, typically in a controlled environment such as a testbed.

\begin{description}
\item[Software setup:] Describe the operating system, any non-standard libraries, including version information, and the hardware environment, including network interfaces, memory size, and graphics cards. For libraries, note if these are not readily available, e.g., due to licensing restrictions. If you used an emulator (e.g., for network links), describe the configuration in detail.

\item[Data input and setup:] What data sources drove the input for your system? What were sources of randomness?

\item[Limitations:] Are you aware of any limitations in your system that may have influenced the measurements, such as performance limitations of the hardware, other experiments sharing the same infrastructure, caches or timing resolution and clock synchronization between systems?

\item[Experiments:] How often did you repeat the experiment? What was the set of parameters you used? (As above, be careful to use unambiguous units and explain if necessary.) It is also good to be aware of common pitfalls that affect the validity of benchmarking results~\cite{benchmarking:crimes1, benchmarking:crimes2} in systems research.

\item[Analysis and presentation:] See \autoref{sec:simulation}.

\item[Data access:] Are any of the traces or raw data available to others? Did you document the log or trace file format? Is it unambiguous which data trace or log correspond to which experiment or measurement? Is the data public or restricted, for instance under non-disclosure agreement (NDA)? Do you anticipate that the data will only be available for a limited time, e.g., because it is a rolling data collection? Consider getting a Digital Object Identifier (DOI) for your data set to make it easy to reference.

\end{description}


\subsection{Real-world Measurements}\label{sec:measurements-real-systems}

Measurements help understand how real systems function. For example, research might measure the current state of deployment of a protocol or feature in the Internet, the characteristics of Internet usage or the behavior of congestion control, security and routing protocols. Measurements can also complement simulations by observing how well a proposed system or protocol functions in the Internet or a real campus or data center network. Measurements can be intra- and inter-domain, measuring the whole Internet, one or more Internet service providers, or a single data-center. Unlike for the previous case, you typically have very limited control over your measurement environment.

\begin{description}
\item[Setup:] Where were your measurement vantage points? For Internet measurement points, what kind of networks were they located in? Do you know the service provider, organization, access technology or geographic location? How did you choose them? For many measurements, the number and location of the measurement vantage (observation) points determines whether the results you obtain are only narrowly or more broadly applicable.

What software did you use to collect the data, e.g., IPFIX~\cite{rfc7011}, Netflow, \texttt{traceroute}, your own mobile application? Did you rely on a public measurement infrastructure, e.g., RIPE Atlas~\cite{bajpai:comst:2015}; Planetlab~\cite{chun:ccr:2003}; etc. Describe the software version and execution environment, such as the operating system and any relevant libraries. What hardware (vendor, model, version or model year) did you use, including any special network interfaces, dedicated flow exporters or special-purpose switches? Do your measurements rely on precise time and how did you ensure clock synchronization both between measurement points and to absolute time?

When running active measurements, characterize your traffic sources. For passive measurements, describe whether you collected all traffic or sampled traffic.

\item[Data collection:] Do the measurements represent a snapshot in time or a longitudinal observation? Justify your sampling period (e.g., a subset of packets versus comprehensive packet capture), the frequency of data collection (e.g., hourly, daily, randomly), and the number of times the data collection has been repeated.

Time and date may influence your results. When was the measurement collected? Be sure to clearly state the timezone. While UTC is generally preferred, in cases where your measurements depend on human diurnal cycles, it may be helpful to capture the local time.

Document all external data sources, such as routing tables, that you collected or that are provided by third parties. If the additional data sources do not describe the same time interval or locations as your collected data, mention this and justify why you consider the data to be applicable. Furthermore, when you measure in an open system, such as the Internet, which is subject to uncontrolled changes, you need to collect and document all relevant metadata (\autoref{sec:documentation}) about the system itself during the measurements. This requires much more planning of the measurements compared to a controlled lab testbed setup where the system aspects are mostly static and can likely be inspected after the measurements have finished. For example, if you work with the Alexa 1M most-popular web site lists, it should be clear which version of the list you actually used~\cite{Alexa-pitfalls}. But even then there is a dynamic mapping of names to addresses using the DNS --- it may matter where, when and how you resolve the names to addresses. If you use a distributed set of vantage points, you will sooner or later need to understand the topology as seen from the perspective of the vantage points. Hence, it is best to collect \texttt{traceroute} data (and if relevant name resolution data) with your measurements as this will be crucial later on to interpret your data set. 

Any missing data needs to be mentioned, particularly data gaps in the collection of measurements caused by operational outages or system maintenance.

\item[Limitations:] Are there limitations that may affect the validity or accuracy of your measurement data or may bias your results?

\item[Analysis and presentation:] See \autoref{sec:simulation}.

\item[Data access:] See \autoref{sec:software-implementation-performance-prototyping}.

\item[Ethics considerations:] Do your measurements implicate potential ethical concerns, in particular those that anybody reproducing your work may need to be aware of? For example, you should document any constraints imposed by institutional review boards or ethics committees. This will also help reviewers judge whether you are complying with general community guidelines~\cite{IMC2007-shared-measurements,Menlo-report}, or those of conferences such as ACM \emph{Internet Measurement Conference} (IMC).
\end{description}
\subsection{Human Subject and Subjective Experiments}\label{sec:human-subject-experiments}

In subjective experiments, participants evaluate the usability or quality of experience (QoE) of a service, functionality, or software. Often, you are testing a hypothesis (``my system works better than the old system'', ``Variable X improves task performance''), which should be formulated ahead of time.

\begin{description}
\item[Setup:] Who were the experimental subjects, e.g., by age brackets, gender, education, and computing skills? Had the subjects taken part in similar experiments before? How did you solicit volunteers? If applicable, note the tracking number for your IRB (Institutional Review Board) or ethics committee approval.

\item[Experiments:] Describe how the experiment was conducted. Were the subjects provided with instructions or just handed your artifact? Were they asked to complete specific tasks? Did the subjects communicate with each other or perform tasks independently?

\item[Limitations:] How did your experiment deviate from ``real life'', e.g., in duration or nature of the task?

\item[Analysis and presentation:] See \autoref{sec:simulation}.

\item[Ethics considerations:] Human subject experiments will likely require approval by an institutional review board (IRB) or ethics panel. You should document key considerations~\cite{Menlo-report, Menlo-report-companion} for protecting human subjects that anybody replicating your study should be aware of and make your IRB filing available to others. (Following the same process during a replication does not relieve the replicator from the duty of seeking approval from an IRB or ethics panel, nor does it guarantee that such approval will be granted.)
\end{description}
\section{Tool Recommendations}\label{sec:tool-recommendations}

Try to use common tools that are widely and readily available. Only develop your own tools if there are no reasonable alternatives. Writing good tools almost always takes longer than you were initially planning for and you will have to maintain those tools for both yourself, other team members and other researchers trying to reproduce your results. We have found the following tools useful in experimental networking research: Document your work in shared lab notebooks using Jupyter. Package your software as containers, using Docker and Kubernetes, or virtual machines (using Vagrant) to avoid dependency hell and ease execution in different environments. For instance, \texttt{ReproZip}~\cite{reprozip} facilitates packaging of experiments by tracking and identifying its dependencies. Use version control (with Git and its ecosystem such as Github and Gitlab) for software and scripts, including scripts for plotting (e.g., based on Python \texttt{matplotlib}, \texttt{R} or \texttt{gnuplot}). You may also want to create software releases by using Git tags. It is always good practice to provide citation credits to authors of the software that you used in the paper.

Many research institutions offer centralized resources for storing experimental data for the long term. Do not store any valuable experimental data on personal devices, laptops or computing devices used for experiments. An experiment may accidentally trash a disk or brick the machine. Your lab will eventually retire old computers and you do not want to have to guess which of the terabytes of data are still valuable. Preservation of digital artifacts is crucial to reproducible research. Preferably, such artifacts can be stored in the digital libraries or at a persistent location that assigns a stable DOI entry. Zenodo~\cite{zenodo} is an open-access repository that can be used to upload and assign DOI to digital artifacts that be referenced in the paper.
\section{Additional Resources} 
\label{sec:resources}

The proceedings of the ACM SIGCOMM 2003 Workshop on Models, Methods and Tools for Reproducible Network~\cite{carle:mometools:2003} and the ACM SIGCOMM 2017 Workshop on Reproducibility~\cite{bonaventure:repducibility:2017} summarize past discussions on this topic. The Stanford University Reproducibility course~\cite{Stanford-systems-course-CCR} is a good example of how students can take published research and attempt to reproduce and document the findings. A list of accepted papers in SIGCOMM-sponsored conferences that released artifacts were recently (2017) surveyed and a compiled list has been made available~\cite{artifact-survey}. The recently established SIGCOMM Artifacts Evaluation Committee carried this initiative forward and applied badges to accepted papers in SIGCOMM-sponsored events in 2018. For instance, CoNEXT 2018 has published the badges of these accepted papers both in the conference proceedings and on the conference web page. Such lists of papers with released artifacts and badged papers can be a good starting to point for students to get started with reproducing published research. Papers by Allman \cite{Allman:Emplirical-Assement} and Reuter et al. \cite{rbcks-trmma-18} discuss the dynamism and heterogeneous nature of the Internet. Other interesting papers highlight pitfalls with IP-address-based geolocation~\cite{geolocation-pitfalls}, the popularity of webpages~\cite{Alexa-pitfalls}, and \texttt{traceroute}~\cite{paristraceroute} that our community has learned from over the past years of empirical research. Other scientific communities are engaging in extensive efforts to improve replicability and reproducibility \cite{rOpenSci,osf}.

\subsection*{Acknowledgments}
The ideas in this paper were developed at the Dagstuhl Seminar \#18412 on ``Encouraging Reproducibility in Scientific Research of the Internet''~\cite{dagstuhl-report-reproducibility} that took place in October 2018. J\"urgen Sch\"onw\"alder and Olivier Bonaventure provided valuable input to this manuscript.

\bibliographystyle{ACM-Reference-Format}
\bibliography{paper}

\end{document}